\documentclass{article}

\PassOptionsToPackage{numbers, sort&compress}{natbib}

\usepackage[final]{neurips_2025_ml4ps}

\usepackage[utf8]{inputenc} 
\usepackage[T1]{fontenc}    
\usepackage{hyperref}       
\usepackage{url}            
\usepackage{booktabs}       
\usepackage{amsfonts}       
\usepackage{nicefrac}       
\usepackage{microtype}      
\usepackage{xcolor}         
\usepackage{bm}
\usepackage{mathtools}      
\usepackage{multirow}
\usepackage{physics}     
\usepackage{amsmath}
\usepackage{amssymb}
\usepackage{graphicx}
\usepackage{enumerate}
\usepackage{enumitem}

\definecolor{hypercolor}{RGB}{174, 60, 60} 
\hypersetup{
  linkcolor  = hypercolor,
  citecolor  = hypercolor,
  urlcolor   = hypercolor,
  colorlinks = true
}

\newcommand{\red}[1]{{\color{red} #1}}

\newcommand{\mvec}[1]{{\bm{#1}}}

\title{High-resolution weak lensing mass mapping from DES-Y3 data using diffusion-based prior}

\author{%
  Supranta S. Boruah\\
  Department of Physics and Astronomy, 
  University of Pennsylvania, \\
  Philadelphia, PA 19103 
  \texttt{supranta@sas.upenn.edu} \\
  \And
  Michael Jacob\\
  Department of Physics and Astronomy, 
  University of Pennsylvania, \\
  Philadelphia, PA 19103 
  \texttt{mgjacob@sas.upenn.edu} \\
  \And
  Bhuvnesh Jain\\
  Department of Physics and Astronomy, 
  University of Pennsylvania, \\
  Philadelphia, PA 19103 
  \texttt{bjain@physics.upenn.edu} \\
  \And 
  Riya Maiya\\
  Department of Physics and Astronomy, 
  University of Pennsylvania, \\
  Philadelphia, PA 19103 
  \texttt{rmaiya@sas.upenn.edu} \\
  \And 
  Raghav Venkataramanan\\
  Department of Physics and Astronomy, 
  University of Pennsylvania, \\
  Philadelphia, PA 19103 
  \texttt{krvenkat@seas.upenn.edu} \\
}

\begin{document}

\maketitle

\begin{abstract}
High-resolution mapping of cosmic mass distribution is essential for a variety of astrophysical applications including understanding cosmic structure formation, and galaxy formation and evolution. However dark matter is not directly observed and therefore we need advanced methods for solving inverse problems to reconstruct the underlying cosmic matter distribution. Here, we train a generative diffusion model and use it in the Diffusion Posterior Sampling (DPS) framework to reconstruct mass maps from Dark Energy Survey-Year 3  (DES-Y3) weak gravitational lensing data at high (1 arcminute) resolution.  We show that the standard DPS results are biased, but they can be easily corrected by scaling the log-likelihood score during the diffusion process, yielding unbiased results with proper uncertainty quantification. The resulting mass maps reveal cosmic structures with enhanced detail, opening the door for improved astrophysical studies using the obtained mass maps.
\end{abstract}

\section{Introduction}\label{sec:intro}

When light from distant galaxies passes near massive structures, it gets bent, subtly distorting the galaxy shapes we observe. This effect, called weak gravitational lensing, allows us to map the invisible dark matter distribution in the Universe without relying on luminous tracers.
These mass maps are fundamental for understanding dark matter structures \cite{Moews2021, Kovacs2022}, constraining cosmological parameters  \citep{Gatti2024_WPH, Jeffrey2025} and understanding the physics of galaxy formation and evolution in different astrophysical environments \cite{Eckert2015, Yang2020}.

Traditional mass mapping approaches face significant limitations in extracting the full potential from modern surveys. The widely-used Kaiser-Squires reconstruction \cite{Kaiser1993} results in highly noisy reconstructions, while Wiener filtering \cite{Wiener1949, DES_Y3_massmapping} assumes Gaussian priors that are inadequate for capturing the highly non-Gaussian nature of late-time cosmic structure. Although lognormal models \cite{Boruah2022, Fiedorowicz2022a, Zhou2024} offer some improvement over Gaussian assumptions, they remain insufficient for the precision requirements of Stage IV surveys \cite{Boruah2024_gansky}. These limitations restrict our ability to utilize small-scale structures from weak lensing mass maps for astrophysical studies.

Machine learning offers a promising path forward to solve the inverse problem of reconstructing cosmic matter distribution. Specifically, we can use generative machine learning models to learn complex, non-Gaussian priors directly from $N$-body simulations and then use this prior to reconstruct the matter distribution in the Universe with a high resolution. Several works have applied generative models to reconstruct dark matter maps from weak lensing data. \cite{Floss2024} used conditional diffusion models for CMB lensing mass maps, learning the posterior directly, while \cite{Remy2023} employed a similar method as us by using score-based diffusion models to learn dark matter map priors, and then used a convolved likelihood approach for the mass map reconstruction. Other generative methods include \cite{Shirasaki2021, Whitney2024}, with \cite{Shirasaki2021} being the only prior work to apply generative machine learning methods to Stage-III survey data to the best of our knowledge. In this paper, we use diffusion models \cite{Ho2020, Song2020} to learn the cosmic matter distribution from $N$-body simulations and then use the Diffusion Posterior Sampling \citep[DPS, ][]{Chung2022_DPS} approach to reconstruct mass maps from Dark Energy Survey Year 3 (DES-Y3) weak gravitational lensing data. 
We have previously demonstrated the application of this method to simulated weak lensing maps in \cite{Boruah2025}.
We show that the standard DPS results are biased, but they can be corrected by using hyperparameters for likelihood scaling. Our method achieves accurate mass map reconstruction 
with well-calibrated uncertainties. The resulting mass maps reveal smaller-scale structures beyond those captured in earlier reconstructions, presenting new opportunities for astrophysical studies.
Our work provides the first diffusion-based reconstruction of weak lensing mass maps from Stage-III surveys with comprehensive validation.

\section{Weak lensing mass mapping using diffusion posterior sampling}

Diffusion models \citep{Song2020, Ho2020} are generative models that progressively add noise to training data until it becomes pure Gaussian noise, then use a neural network to reverse this process. The forward diffusion process gradually adds noise over timesteps, while the reverse process uses the gradient of the log-probability density, $\nabla_{\mvec{x}_t} \log p_t(\mvec{x}_t)$ (the `score' function). A neural network estimates this score function: $\mvec{s}_{\mvec{\theta}}(\mvec{x}_t) \approx \nabla_{\mvec{x}_t} \log p_t(\mvec{x}_t)$. Once trained, the model generates samples by starting from Gaussian noise and iteratively denoising. 

\subsection{Diffusion Posterior Sampling and likelihood scaling}

\begin{figure}[h]
    \centering
    \begin{minipage}{0.65\textwidth}
        \centering
        \includegraphics[width=\textwidth]{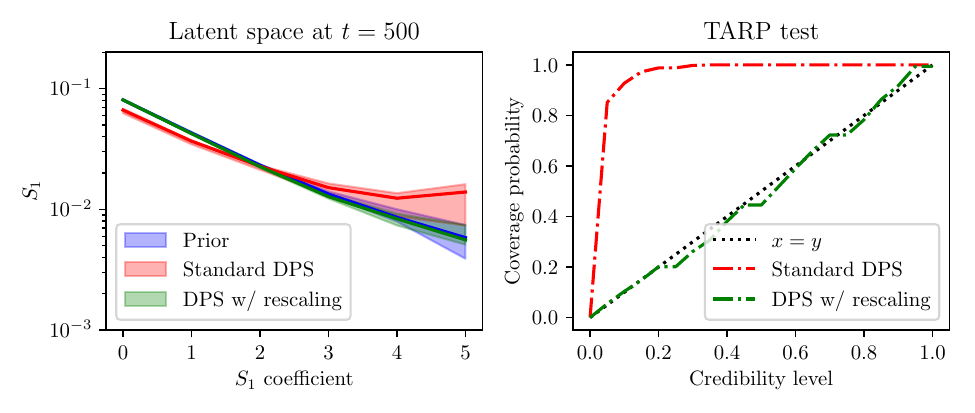}
    \end{minipage}
    \hfill
    \begin{minipage}{0.34\textwidth}
        \caption{\textit{Left:} Standard DPS (\textit{red}) makes the latent maps go out-of-distribution. We resolve this with our likelihood scaling method (\textit{green}). \textit{Right:} TARP coverage test showing well-calibrated uncertainty quantification (following the $x=y$ line) with likelihood scaling, whereas the standard DPS gives biased results. 
        }
        \label{fig:ood_tarp}
    \end{minipage}
\end{figure}

For weak lensing mass mapping, we reconstruct the convergence field $\mvec{\kappa}$,  from noisy shear observations $\mvec{D}$. To sample from the posterior, we need the gradient of the log-posterior \cite{Anderson1982} which is given using the Bayes' theorem as:

\begin{equation}\label{eqn:posterior_score}
\nabla_{\mvec{\kappa}_t} \log p(\mvec{\kappa}_t|\mvec{D}) = 
\nabla_{\mvec{\kappa}_t} \log p(\mvec{D}|\mvec{\kappa}_t) + \nabla_{\mvec{\kappa}_t} \log p(\mvec{\kappa}_t),
\end{equation}
where $\mvec{\kappa}_t$ is the latent convergence map at timestep $t$. The likelihood gradient $\nabla_{\mvec{\kappa}_t} \log p(\mvec{D}|\mvec{\kappa}_t)$ is intractable \cite{Chung2022_DPS}. Diffusion Posterior Sampling (DPS) \citep{Chung2022_DPS} approximates this using Tweedie's formula \citep{Robbins1992_Tweedie, Kim2021_Tweedie}, evaluating the likelihood at the posterior mean which we denote as $\hat{\mvec{\kappa}}_{0|t}$. We refer the readers to \cite{Chung2022_DPS} for details of DPS. DPS produces biased inference due to accumulated errors during early reverse diffusion \cite{daps}. We address this with likelihood scaling, multiplying the likelihood gradient by a time-dependent factor $A(t)$:

\begin{equation}
\nabla_{\mvec{\kappa}_t} \log p(\mvec{\kappa}_t|\mvec{D}) \approx A(t) \cdot \nabla_{\mvec{\kappa}_t} \log p(\mvec{D}|\hat{\mvec{\kappa}}_{0|t}) + \mvec{s}_{\mvec{\theta}}(\mvec{\kappa}_t), 
\end{equation}
where $A(t) = \text{Sigmoid}((\bar{t}-t)/\sigma_t)$ with $\bar{t} = 500$ and $\sigma_t = 20$. This down-weights the likelihood at early timesteps when samples are prone to going out-of-distribution (Figure \ref{fig:ood_tarp}, left), while maintaining full weighting at final steps. 
To determine $\bar{t}$ and $\sigma_t$, we run the TARP test (section 3.1) and search for hyperparameters that provide a good coverage test. The prior score is estimated using a neural network as $\mvec{s}_{\mvec{\theta}}(\mvec{\kappa}_t)$. The details of the neural network and its training can be found in \cite{Boruah2025}.

\subsection{Training data and likelihood}

We train our diffusion model on 50 full-sky ray-traced convergence maps from {\sc pkdgrav3} simulations \cite{Potter2016}, creating 4 tomographic maps matching DES-Y3 source redshift distributions \cite{DESY3_nz}. 
These simulations are all produced at a fixed cosmology and the diffusion model learns the dark matter distributions in these simulations as the prior.
From each of these simulations, we extract 2,146 flat $256\times256$ pixel maps per full-sky map at 1 arcminute resolution (resulting in a total of $>100k$ flat maps).
We assume Gaussian likelihood for shear observations with independent shape noise per pixel:
\begin{equation}\label{eqn:wl_likelihood}
    p(\mathbf{D}|\boldsymbol{\kappa}) \propto \prod_{\alpha=1}^{2} \prod_{i=1}^{N_{\text{pix}}} \exp\left[-\frac{(\gamma_{\text{obs},\alpha}^i - \gamma_\alpha^i(\boldsymbol{\kappa}))^2}{2\sigma_{\epsilon,i}^2}\right]
\end{equation}
where $\gamma_{\text{obs},\alpha}^i$ are observed ellipticity, $\gamma_\alpha^i(\boldsymbol{\kappa})$ is shear calculated from the convergence field, and $\sigma_{\epsilon,i}$ is shape noise in the $i$-th pixel. While DES-Y3 shape noise is non-Gaussian \cite{DES2024_SC}, our framework can incorporate non-Gaussian likelihoods without retraining the diffusion model. We will assess the impact of non-Gaussian noise in a future work.

\section{Results}\label{sec:results}

\subsection{Validation on simulation}\label{ssec:dps_sims}

We validate our method using mock DES-Y3 observations generated from simulated convergence maps. We create realistic mock shear data that includes a survey mask and Gaussian shape noise with $\sigma_{\epsilon} = 0.26$ per galaxy, using number densities corresponding to the DES-Y3 survey. Pixels outside the survey footprint are masked during likelihood evaluation Equation \eqref{eqn:wl_likelihood}. 
Sampling 1 posterior map requires around 1 minute on a NVIDIA A40 GPU.

We demonstrate that the standard DPS method produces biases when early steps go out of distribution (Figure \ref{fig:ood_tarp}, left panel). However, our modified DPS algorithm with likelihood scaling successfully corrects this bias and reconstructs the underlying mass distribution.

To assess the uncertainty quantification of our method, we employ the Tests of Accuracy with Random Points \citep[TARP, ][]{TARP_test2023} framework. TARP provides a sampling-based calibration test that evaluates whether generated posterior samples accurately represent the true marginal posterior distribution. Through hyperparameter tuning of the likelihood scaling, we achieve well-calibrated posteriors with accurate uncertainty quantification (right panel of Figure \ref{fig:ood_tarp}). We refer the readers to \cite{TARP_test2023} for more details on the TARP test.

\begin{figure*}
    \centering
    \includegraphics[width=\linewidth]{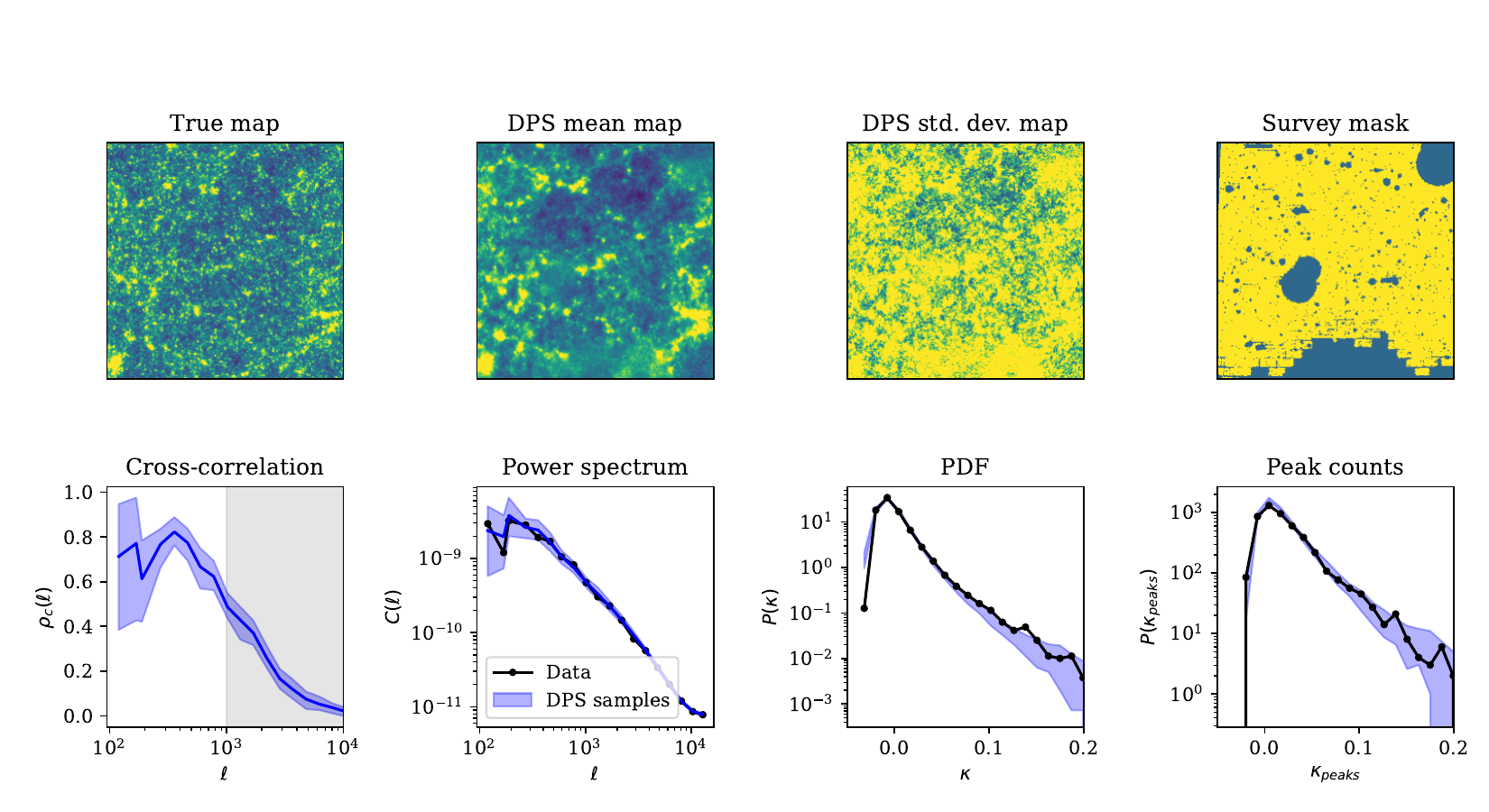}
    \caption{
    Validation of our method on simulations. \textbf{Top row}: True underlying mass map (left), mean of reconstructed mass map samples (center left), standard deviation of reconstructed samples (center right), and survey mask (right). \textbf{Bottom row}: Cross-correlation between reconstructed samples and true map (left), followed by comparisons of power spectrum, PDF, and peak counts between reconstructed samples (blue) and true map (black).
    }
    \label{fig:sim_validation}
\end{figure*}

Figure \ref{fig:sim_validation} demonstrates the performance of our method on mock simulations. Visual inspection of the maps confirms that we successfully reconstruct prominent dark matter features. We compute the cross-correlation between the reconstructed maps and the true map, showing that we correctly reconstruct structure on scales up to $\ell \lesssim 1000$. This represents an improvement of almost a factor of 5 over state-of-the-art mass mapping methods. For example, \cite{Boruah2024_massmapping} achieves $\rho_c(\ell)\approx0.5$ at $\ell \sim 200$, whereas our maps achieve the same correlation level at $\ell \sim 1000$. We further validate our method's performance by computing the power spectrum and other non-Gaussian summary statistics. Our reconstructed maps show excellent agreement with the true underlying maps across all these statistics.

\subsection{Mass map reconstruction of DES-Y3 data}\label{ssec:dps_desy3}

\begin{figure*}
    \centering
    \includegraphics[width=\linewidth]{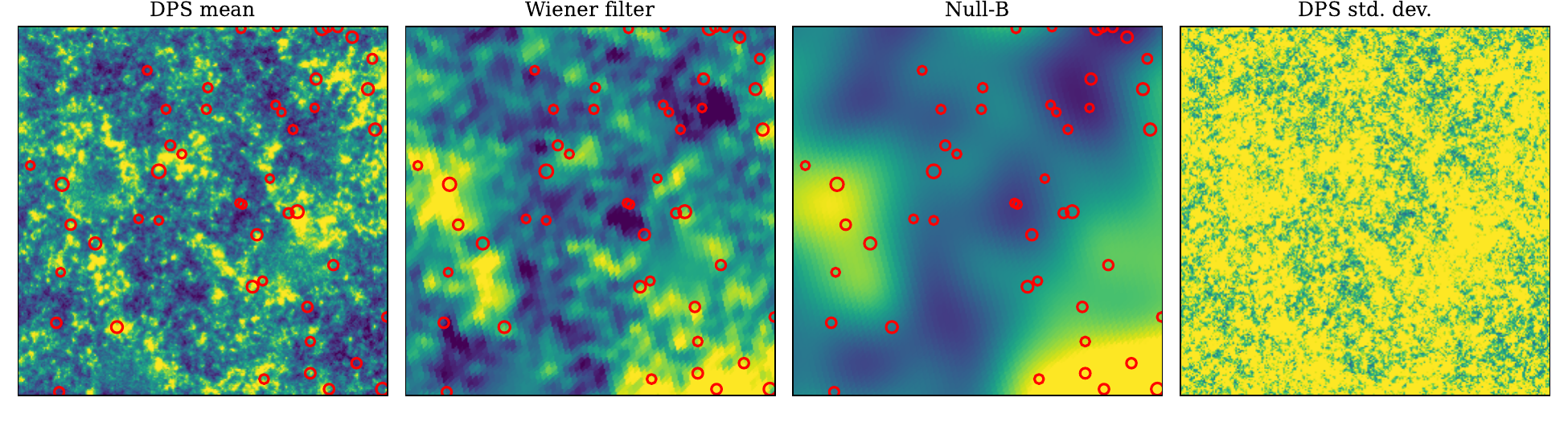}
    \caption{
    Comparison of the reconstructed maps from the DES-Y3 data in a $4.26^{\circ} \times 4.26^{\circ}$ representative patch. The left panel shows the mean of the DPS reconstructions, the center left panel shows the Weiner filtered reconstruction, and the third panel shows the Null-B reconstruction. The right panel shows the standard deviations computed from the DPS reconstructions. DES-Y3 RedMapper clusters are overplotted with red circles.  
    }
    \label{fig:map_plot_des_y3}
\end{figure*}

After validating our method on simulated convergence maps, we apply it to DES-Y3 shear data. We created pixelized maps of mean galaxy ellipticity at 1 arcminute resolution. While we present in this section our results for only one of the patches, we have used our method to create the mass maps for the full DES-Y3 footprint by dividing the data in 410 patches of the same size. Our DPS maps are reconstructed independently on each flat patch. Figure \ref{fig:map_plot_des_y3} shows the reconstruction for one representative patch. We compare our DPS reconstruction (left) to Wiener filter and Null-B maps of the same data produced in \citep{DES_Y3_massmapping}. From the figure, we see that: {\it i)} while our maps show overdensities and underdensities in the same sky regions as the other maps, they reconstruct structures on much smaller scales, and {\it ii)} DES-Y3 RedMaPPer clusters (red circles) correlate strongly with high-convergence regions in our maps.

We plan to perform additional validation of our reconstructed maps by cross-correlating them with external datasets such as galaxy catalogs and Compton $y$-maps. The enhanced small-scale reconstructions will provide a powerful tool for probing galaxy formation and evolution at unprecedented resolution. The rightmost panel in Figure \ref{fig:map_plot_des_y3} shows the pixel-wise uncertainty in our reconstruction, which is essential for downstream analyses.
\section{Conclusion}
In this paper, we present a diffusion-based reconstruction of mass maps from DES-Y3 weak lensing data. This represents one of the first applications of generative AI-based mass map reconstruction to Stage-III survey data. By applying our diffusion-based approach to DES-Y3, we have produced high-resolution mass maps with robust uncertainty quantification. Our maps improve the resolution of mass map reconstruction by more than a factor of 5 compared to previous methods applied on the same data.

These high-resolution maps reveal significantly more small-scale structures than earlier reconstructions, opening new avenues for astrophysical studies. This enhanced resolution enables novel scientific applications including the identification of filaments and voids, and investigations of galaxy formation and evolution in different dark matter environments.

Future work will carefully examine the impact of various assumptions made in this study. The plug-and-play nature of DPS makes it straightforward to incorporate these improvements without retraining our diffusion model. One limitation of our current method is the need to carefully tune likelihood scaling hyperparameters. We plan to implement more robust methods for solving inverse problems with diffusion models to address this limitation \citep[e.g,][]{Tolooshams2025, Karan2025}.
Another potential limitation is that because our simulations cannot capture every aspect of the real data, the observed data may not lie within the training distribution. Understanding how such out-of-distribution (OoD) data impact the inference remains an open question. We expect these effects to introduce biases, and we need to explore strategies to quantify and reduce them is left for future work. Finally, we have not compared our results with other ML-based methods, as DES-Y3 mass maps generated using such approaches are not publicly available. A detailed comparison across different ML-based weak-lensing mass-mapping techniques is planned for future work.

\section*{Acknowledgements}
We thank Gary Bernstein, Marco Gatti, Matt Ho, Mike Jarvis, Laurence Perreault-Levasseur and Kunhao Zhong for useful discussions and the anonymous referees for useful comments. S.S.B and B.J. are partially supported by the US
Department of Energy grant DE-SC0007901 and by NASA funds for the Open Universe project. 
\bibliography{DPS_WL}



\end{document}